\documentstyle[12pt,aaspp]{article} 
\begin{document} 
 
\title{A STUDY INTO THE FEASIBILITY OF OBTAINING FINITE SOURCE SIZES 
FROM MACHO-TYPE MICROLENSING LIGHT CURVES} 
\author{Eric W. Peng} 
\affil{Princeton University Observatory, Princeton, NJ 08544--1001} 
\affil{\tt e-mail: ericpeng@astro.princeton.edu} 
 
\begin{abstract} 
Recent discussion of the effects of finite source size on high 
magnification microlensing events due to MACHOs motivates a study 
into the feasibility of observing such effects and extracting the source 
radius.  Simulated observations are generated by adding Gaussian error 
to points sampled on theoretical microlensing light curves for a limb 
darkened, extended source.  These simulated data sets are fitted in an 
attempt to see how well the fits extract the radius of the source. 
The source size can be fitted with reasonable accuracy 
only if the impact parameter of the event, $p$, is less than the 
stellar radius, $R_{\star}$.  It is possible to distinguish ``crossing'' 
events, ones where $p<R_{\star}$, from ``non--crossing'' events if the 
light curve is 
well sampled around the peak and photometric error is small --- 
i.e.~$\geq 3$ observations while the lens transits the disk of the 
source, and 
$\sigma_{phot}<0.08\;{\rm mag}$.  These requirements are just within the 
reach of current observational programs; 
the use of an early-warning system and multiple observing sites should 
increase the likelihood that  $R_{\star}$ can be fitted.  The programs 
used to simulate and fit data can be obtained via anonymous {\tt ftp}. 
 
\end{abstract} 
 

\section{INTRODUCTION} 
 
Paczy\'nski (1986) noted that 
if the dark matter constituents of the halo are massive enough, 
then the optical depth to microlensing is on the order of $10^{-6}$. 
In the past few years, searches for gravitational microlensing events 
in the LMC and the Galactic Bulge have discovered many plausible 
candidates through long-term monitoring of millions of stars 
(OGLE: Optical Gravitational Lensing Experiment, 
Udalski et al. 
1993; EROS: Experience de Recherche d'Objets Sombres, 
Aubourg et al. 1993; MACHO: Massive Compact Halo Objects 
collaboration, Alcock et al. 1993; DUO: Disk Unseen Objects program, 
Alard et al. 1995). 
All these projects share the goal of trying to measure the 
amount of dark matter in the Galactic halo that is in the form of 
massive compact halo objects (MACHOs). 
 
This study is motivated by recent papers that show the shape of 
the light curve around the peak to be considerably different from the 
one predicted for a point source in events where the impact parameter 
is less than or equal to the source radius. 
(Witt~\&~Mao 1994; 
Nemiroff~\&~Wickramasinghe 1994; Gould 1994).  The ability to use the light 
curve of an event to determine the source size, $R_{\star}/R_E$, 
where $R_E$ is the Einstein radius, is valuable because we 
already have an independent estimate of $R_{\star}$ using spectral type. 
This allows us to use the shape of the light curve to obtain a direct 
estimate of
the value of $R_E$, which 
helps to obtain a more accurate estimate of the mass of 
the lens. 
The traditional way of estimating $R_E$ requires assumptions of 
the statistical measures of the relative velocities. 
Also, knowing the values of $R_E$ and $t_0$, the time 
it takes for the lens to traverse a distance $R_E$ relative to the 
source 
in the deflection plane, gives us the transverse velocity. 

The approach taken in this paper is limited to single-lens microlensing.  
Witt (1995) predicts that 3\% of microlensing events caused by lenses
in the Galactic bulge should show noticeable effect due to the finite
source size.
Deviations from the point source light curve due to effects such as 
parallax and binary lenses were ignored for the purposes of this study.  
Parallax effects have been observed, but occur on time scales much 
larger than those on which finite source size effects occur 
(Alcock et al. 1995).  Binary lens events, while not 
insignificant, should only make up $\sim 10\%$ of microlensing 
candidates toward the Galactic bulge (Mao \& Paczy\'{n}ski 1991).  
However, in the future, the technique used in this 
study may also be used to investigate these types of deviations.

While it is important to identify the possible effects of finite source 
size on the light curve, it is equally important to test the feasibility 
of observing such effects. 
After a systematic study of generated light curves for single-lens 
events, this paper 
determines a lower limit on the sampling rate and photometric accuracy 
of observations needed to 
reliably extract $R_{\star}$ from an event. 
The outline of the 
paper is as follows:  in \S 2, 
the general method for generating light curves and the approach to 
feasibility testing are discussed; in \S 3, the results are presented 
and recommendations for observing programs are made. 
 
\section{METHOD} 
 
The shape of the light curve for a gravitationally 
lensed extended source can be fully described by 
five parameters: $p$, the impact parameter, which is the angular 
separation between the lens and the unlensed position of 
the source at the time of maximum magnification (from now on 
given in units of $R_E$); $R_{\star}$, the source radius 
(also given in units of $R_E$); $t_0$, the time it takes for the 
lens to traverse a distance $R_E$ relative 
to the source in the deflection plane; $t_{max}$, the time of 
maximum magnification; and $mag_0$, the unlensed, or baseline 
magnitude of the source. 
 
The Einstein ring radius, the distance an image would appear from 
the line of sight if the lens and source were exactly aligned is given 
by 
\begin{equation} 
R^2_E = \frac{4GMD}{c^2},\; D\equiv \frac{D_d(D_s-D_d)}{D_s} 
\end{equation} 
where $D_d$ and $D_s$ are the distances to the lens of mass $M$ and 
to the source, respectively. 
 
The expression for the observed magnitude of a disk of constant surface 
brightness as a function 
of position along the trajectory of the lens 
is derived by Witt~\&~Mao~(1994).  In this paper, however, all sources 
are considered to be limb darkened disks with limb darkening parameter 
$0.6$, using the formulation of Aller (1953, p. 207). 
 
When the source and lens move relative to each other in the deflection 
plane, the magnification varies with time.  An example of such 
a light curve is depicted in Figure~1. 
Each light curve is labeled with the impact parameter of 
the source trajectory.  Time is given in units of $t_0$ and $R_{\star}$ 
for all curves is $0.055$ $R_E$.  Parts of the computer code used to 
generate these light curves were obtained from Bohdan Paczy\'nski. 
 
The ability to simulate observations not only gives one a large 
number of sample data sets on which to test potential observational 
programs, 
but it also facilitates error analysis by allowing one to compare the 
values extracted by a fitting routine with the ``true'' input 
values.  In this model, input parameters were chosen and the 
theoretical light curve was evenly sampled within each daily 
observing period, with sampling frequency $n$. 
In theory, the length of this daily 
observing period can range from 1--24 hours. 
The photometric errors were simulated by adding a random 
Gaussian error in magnitude to each one of these points. 
The photometric error for a given observation, 
$\sigma_{phot}$, is a function of magnitude and is given by 
\begin{equation} 
\sigma_{phot}(mag) = \sigma_0 10^{(mag_0 - mag)/3.5},\; 
\sigma_0=\sigma_{phot}(mag_0) 
\end{equation} 
(Udalski et al. 1994a).  Values for $\sigma_0$ range from $0.01 \;{\rm 
mag}$ to $0.25 \;{\rm mag}$. 
This dependence was derived from an empirical fit to median OGLE error 
at different magnitudes.  In the future, with better statistics, we will 
be able to use error histograms to empirically fit the error 
distribution, and thereby avoid making assumptions of the Gaussian 
nature of the error. 
 
To test for fitting accuracy, a simulated data set is generated 
with specific, known parameters. 
Then, a best-fit microlensing light curve is obtained with a program 
that 
uses the Levenberg--Marquardt method for nonlinear fits in 
multiple dimensions to minimize $\chi^2$ (Press et al. 1992). 
Figure~2 shows a set of simulated observations and the best-fit 
light curve.  The {\tt rms} photometric error at the baseline magnitude, 
$\sigma_0$, 
was taken to be $0.1 \;{\rm mag}$. 
 
Many different data sets were created and fitted with the goal of 
investigating error in the extracted source radius as a function 
of $p$, $\sigma_0$, $n$, and the length of the daily observing period. 
There are a large number of plausible lensing configurations and 
it would be extremely time-consuming and computer-intensive to 
examine them all.  It was the goal of this study to examine a few 
extreme cases and a few pertinent ones so that trends in error would 
be evident, and so that rough, 
conservative guidelines could be suggested for observing programs. 
Those interested in testing different cases can access via anonymous 
{\tt ftp} 
the original C code used to simulate and fit observations at 
astro.princeton.edu.  After login change directory to bp/finite. 
The read.me file contains the information about all other files, their 
names and sizes. 
 
\section{RESULTS AND DISCUSSION} 
 
In the following discussion, the quantity $R_{\star,fit}$ is used to 
describe the fitted value for $R_{\star}$; the value $p_{fit}$ 
describes the fitted value for $p$. 
Figure~3a plots the percent error in $R_{\star ,fit}$ 
against $p_{fit}$.  Every point represents the fit to 
simulated observations of an event with random $p$.  The fitted 
value of the impact parameter is plotted on the x--axis (as opposed to 
$p$)  
because that is what is measured in real data sets.  Each 
extracted value of $R_{\star}$ has an error $(R_{\star,fit}-R_{\star})$. 
With the exception of $p$, all the input parameters are identical 
for every data set.  These input values 
are: $M_{lens}=0.1\;{\rm M_{\sun}}$, $D_d=8\;{\rm kpc}$, $D_s=9\;{\rm 
kpc}$, 
$mag_0=17.0\;{\rm mag}$, $R_{\star}=10\;{\rm R_{\sun}}$, 
$V=100\;{\rm km \;s^{-1}}$, $n=2$, $\sigma_0=0.0222\;{\rm mag}$, 
and 24--hour coverage is assumed.   
 
These values give a $t_0$ of 
roughly 15 days, an $R_{\star}\sim 0.055\;{\rm R_E}$, and represent a 
plausible lensing geometry if one is looking toward the Galactic 
Bulge.  This particular lensing configuration was chosen to illustrate 
the results determined from this study.  Events simulated had source 
radii that ranged from 0.01--1 ${\rm R_E}$, and $t_0$ that ranged from 
1--30 days. 
 
The value for the baseline photometric error, $\sigma_0$, in this and  
all simulations done in this study, 
was taken from an analysis of OGLE photometry by Udalski et al. (1994). 
The points represented by ``x's'' are ``crossing'' event simulations, 
ones where $p<R_{\star}$ and the lens actually crosses the stellar disk. 
Circles signify ``non--crossing'' event simulations, ones where 
$p>R_{\star}$, with the radius of each circle proportional to 
$p-R_{\star}$.  Witt~\&~Mao (1994) have shown that very little 
differentiates light curves for point source microlensing and 
extended source microlensing if the impact parameter is 
larger than the source radius.  Error plots, such as 
Figure~3a, confirm that it is 
difficult to extract the source radius from a light curve if 
the lens does not cross the disk of the source. 
 
Figure~3a also illustrates how photometric error quickly 
blurs any information on the source radius when $p>R_{\star}$. 
On the other hand, for ``crossing'' events, fitting is 
fairly reliable.  The question which follows is: for real observations, 
where we do not know nature's input values, can we distinguish 
``crossing'' data sets from ``non--crossing'' ones? 
The answer to this question is yes, if our observations are accurate 
enough. 
 
The dashed line in Figure 3a is the line on which $p_{fit}=R_{\star 
,fit}$. 
All points to the left of this line represent simulations in which 
$p_{fit} < R_{\star ,fit}$, i.e. apparent crossing events.  All points 
to the right of this line are simulations in which  
$p_{fit} > R_{\star ,fit}$; apparent non--crossing events.  If 
all apparent crossings correspond to true crossings, i.e. all points to 
the  
left of the line are ``x's'', then we are able, from the fitted 
parameters alone, to distinguish true crossings from  
true non--crossings.  Figure 3a shows a scenario in which this is true. 
Likewise, all apparent non--crossing events (right of line)  
correspond to true non--crossing events (circles). 
 
Figure 3b displays the linear relationship between ${\Delta} p$, 
the error in $p_{fit}$, and ${\Delta} R$, the error in $R_{\star ,fit}$. 
As in Figure~3a, every point represents 
simulated observations of an event with a random $p$ and all other  
input parameters held constant. 
We would expect errors in $R_{\star,fit}$ and errors in $p_{fit}$ to be 
coupled because the light curve is most sensitive to both 
of these parameters around the time of maximum magnification. 
Hence, the greater the error in the fitted impact parameter, 
the greater the error in the fitted stellar radius needs to be in order 
fit the points on the light curve. 
The slope of this relationship is a function of the photometric 
errors and the sampling.  Larger errors and fewer observations 
cause the slope to increase.  If this  
``error slope'', $\Delta R_{\star}/ \Delta p$, has  
a value less than one, 
i.e. less than that of the dashed line in Figure~3a, then  
the relation between $p$ and $R_{\star}$ 
will be the same for both input and extracted values; meaning that 
\[ p<R_{\star} \Longleftrightarrow p_{fit}<R_{\star ,fit}.\] 
As a result, we should be able 
to distinguish between ``crossing'' events and ``non--crossing'' events 
using just the fitted values. 
 
The error slope depends on many factors.  For any given lensing 
geometry, the most important ones are: 
the baseline photometric error, $\sigma_0$, and the observational 
coverage, 
determined by both the sampling frequency, $n$, and the length of the 
daily observing period. 
 
\subsection{Photometric Error} 
 
Assuming that there are an adequate number of evenly 
spaced observations on a light curve, the photometric error is 
the quantity that defines the error slope. 
The $\chi^2$ surface as a function of 
both $p$ and $R_{\star}$ possesses a valley in which there is 
a global minimum.  The location 
of this minimum in the valley is very sensitive to errors in 
photometry; this can cause it to move significantly. 
Error slopes were determined using least squares fitting in plots  
like Figure~3b.  Only points for which $\Delta R_{\star ,fit} >0$ were 
fitted. 
This is because: 1) the fact that $R_{\star ,fit}$ cannot be negative 
skews 
the slope for points where $\Delta R_{\star ,fit}<0$, and 2) These 
points  
will never cross into the  
$p_{fit}<R_{\star ,fit}$ regime, as exhibited by the lower half of 
Figure 3a. 
The trend derived from this rough analysis shows that  
for $\sigma_0>0.08{\rm mag}$, 
the error slope is greater than one and grows slowly with increasing 
$\sigma_0$; for $\sigma_0<0.08\;{\rm mag}$, the error slope 
decreases rapidly for decreasing $\sigma_0$. 
The value $\sigma_0=0.08\;{\rm mag}$ can be used as a rough upper limit 
on the 
photometric error.   
 
However, each lensing configuration is different 
and some may be more tolerant of photometric errors than others. 
Since $\sigma_0$ is the baseline photometric error, and the part of  
the light curve with which we are concerned is the peak, the error 
slope will also be a function of lensing amplitude --- photometric error 
decreases as the source brightens.  In events where the source radius  
is large ($R_{\star} > 0.5\;{\rm R_E}$), the total magnification is 
significantly lower than 
for the point source case, and the  
resulting photmoetric errors at the peak are relatively large.   
However, in these events, the effect of the finite disk is  
so pronounced that overall, it is still possible to obtain a reasonable 
fit the source radius provided that there are an adequate number of 
observations during the crossing of the disk by the lens.   
 
\subsection{Sampling Frequency} 
 
Even with dense sampling, errors in photometry still prevent fitting 
from being perfect.  Sparse 
coverage of an event can cause a fitting routine to wander aimlessly,  
not possessing enough information to determine the global minimum. 
Frequent and evenly spaced measurements around the time of 
the peak magnification are essential for good convergence of 
the fitting routine. 
 
Experiments with the fitting of simulated data sets have shown that some 
lensing configurations will be more forgiving toward sparse sampling 
than others. The smaller the impact parameter and the larger 
the source radius in units of $R_E$, the easier it is to ``resolve'' 
the disk.  The most important result obtained from these simulations 
is that regardless of 
how high the sampling frequency is on other parts of the light curve, 
there is very little hope of fitting the source radius without 
at least one observation while the lens is transiting the disk of the 
source.   
 
A conservative estimate requires at least three observations during  
the disk crossing in order to have an error slope less than one.   
The timing of the observations is most important for a good fit. 
A single observation exactly at the time of maximum magnification  
will often result in a good extraction of the source radius.   
In a sense, it is in an effort to obtain this one point that we  
must take as many observations as possible around the time of maximum. 
Multiple observations also reduce the effect of photometric errors and 
increase the certainty of the fit. 
 
\subsection{Observational Programs} 
 
Since the time it takes for a lens to cross the disk of a source is 
usually on the order of one day (and that is for $p=0$), multi--site 
coverage, while not an absolute necessity, would greatly increase 
the chances of resolving a stellar disk. 
If we assume 
that there are only 8 hours a night during which a group can observe 
at one site, then the disk crossing of an event with 
$t_0\sim 4\;{\rm days}$ and $R_{\star}=0.05\;{\rm R_E}$ has 
a window of only $5$ hours, and can be missed 
entirely if timing is less 
than fortuitous.  Moreover, the light curves of short $t_0$ events are 
more prone to exhibit effects due to finite source size because they are 
likely to involve small lensing masses, which implies a small $R_E$ and 
a large $R_{\star}/{R_E}$. 
Early--warning systems, like the ones currently used by the MACHO and 
OGLE groups (Udalski et al. 1994b), allow resources to be  
concentrated on a single event and 
will increase the density of photometry around the time of maximum 
magnification.  Multi--site monitoring of microlensing events, with
telescopes in Australia, South Africa, and Chile, has already
been initiated by the PLANET collaboration (Probing Lensing Anomalies
NETwork; Albrow et al. 1995).
 
To a limited extent, good photometry can compensate for less than 
ideal observational coverage and vice versa.  However, good quality 
in both respects is needed to increase the chances that observations of 
the light curve of a ``crossing'' event will still contain information 
on the source size. 
Nemiroff \& Wickramasinghe (1994) emphasize the importance of having 
enough photometry in order to resolve inflection points in the light 
curve.  However, this demanding criteria need not be met since the 
source size can be fit directly as a free parameter, a technique which 
requires a much lower sampling rate. 
 
The error in $R_{\star,fit}$ shows no discernible trend as 
a function of $t_0$, $t_{max}$, or $mag_0$. 
 
Time resolution and coverage are currently the limiting factors 
in microlensing observations.  Photometric errors in OGLE observations 
are already within the prescribed limits.  Present observational 
programs that are fortunate enough to detect an extremely high  
magnification event in the early stages will most likely be able to  
determine  whether or not it is a ``crossing'' event. 
For shorter $t_0$ events, the ability to take 
observations from more than one site coupled with an early--warning 
system greatly increases the chances of resolving stellar disks.

\acknowledgments 
 
I am grateful to Bohdan Paczy\'nski for his insightful advice and 
enthusiastic support.  I thank Hans Witt, Ralph Wijers, and Raja 
Guhathakurta for helpful comments and discussions.  I also thank 
the referee for useful suggestions.
Krzysztof Stanek, Guohong Xu, and Rongsheng Xu all provided valuable 
assistance with logistics.

\begin{figure} 
\begin{center} 
{\bf FIGURE CAPTIONS} 
\end{center} 
 
\caption{Theoretical light curves for microlensing of an extended, limb 
darkened source.  Shown here are four light curves 
corresponding to 
events with impact parameters $p/R_E=0.0,\;0.055,\;0.2,\;0.5$. 
The time scale $t_0$ is defined as the time it 
takes the lens to move a distance $R_E$ 
with respect to the source, where $R_E$ is the Einstein ring radius. 
The other input values for these curves are: 
$M_{lens}=0.1M_{\sun}$, $D_d=8\;{\rm kpc}$, $D_s=9\;{\rm kpc}$, 
$R_{\star}=10{R_{\sun}}$, $V=100\;{\rm km\,s^{-1}}$.  This gives 
an $R_{\star}/R_E = 0.055$. The $p=0$ event is a ``crossing'' event; 
the $p=0.055$ is an event wher ethe lens grazes the edge of the source; 
the other two are ``non--crossing'' events.} 
 
\caption{Simulated observations and the best--fit light curve.  These 
observations were simulated by sampling a theoretical light curve, like 
the ones shown in Figure~1, and adding a random gaussian error in 
magnitude. 
Here, $\sigma_{phot}$ varies with magnitude with 
$\sigma_0=0.1\;{\rm mag}$. 
The curve is the best-fit light curve obtained using a 
Levenberg--Marquardt 
$\chi^2$ minimization routine.  It is useful to compare this best-fit 
curve to the original theoretical curve to study the errors involved 
in observation and fitting.} 

\end{figure}
\begin{figure}

\caption{a) The percent error in $R_{\star, fit}$ against $p_{fit}$.   
Error in the fitted source size is small if $p<R_{\star}$. 
When $p>R_{\star}$, errors rapidly increase. 
Every point represents a
simulated observation for an event with a random $p$ and all other 
parameters held constant.  Points represented by ``x's'' are 
``crossing'' event simulations.  Circles signify ``non--crossing'' 
event simulations with the radius of the circle proportional to
$p-R_{\star}$.  In addition to the input 
parameters used in Figure~1, $mag_0=17.0\;{\rm mag}$, $n=2$, 
$\sigma_0=0.0222\;{\rm mag}$, 
and 24--hour coverage is assumed.   
The dashed line is the line on which $p_{fit}=R_{\star ,fit}$. 
All points to the left of this line represent simulations in which 
$p_{fit} < R_{\star ,fit}$, i.e. apparent crossing events.  All points 
to the right of this line are simulations in which  
$p_{fit} > R_{\star ,fit}$; apparent non--crossing events.  If 
all apparent crossings correspond to true crossings, i.e. all points to 
the  left of the line are ``x's'', then we are able, from the fitted 
parameters alone, to distinguish true crossings from  
true non--crossings.  This figure exhibits a scenario in which this is 
true. 
All ``crossing'' events were fitted with relatively high accuracy. 
b) The percent error in $R_{\star,fit}$ against the  
error in $p_{fit}$, ${\Delta} p$. 
The relationship between ${\Delta} R_{\star}$, $R_{\star,fit}-
R_{\star}$,  
and ${\Delta} p$ is linear for ${\Delta} R_{\star}$,${\Delta} p > 0$. 
The slope of this relationship is a function of the photometric 
errors and the sampling.  Larger errors and fewer observations 
cause the slope to increase.  If this  
``error slope'', $\Delta R_{\star}/ \Delta p$, has  
a value less than one, 
i.e. less than that of the dashed line in Figure~3a, then apparent 
crossings will correspond to true crossings.}
\end{figure} 

\newpage 
\begin{references} 

\reference Alard, C. 1995, in Proc. IAU Symp. 173 (Eds. C. S. Kochanek,
J. N. Hewitt), Kluwer Academic Publishers, p. 215
\reference Albrow, M. et al. 1995, in Proc. IAU Symp. 173 (Eds. 
C. S. Kochanek, J. N. Hewitt), Kluwer Academic Publishers, p. 227
\reference Alcock, C. A. et al. 1993, Nature 365, 621 
\reference Alcock, C. A. et al. 1995, ApJ, submitted
\reference Aller, L. H. 1953, Astrophysics. The Atmospheres of the Sun 
and Stars, The Ronald Press Co., New York 
\reference Aubourg, E. et al. 1993, Nature, 365, 623 
\reference Gould, A. 1994, ApJ, 421, L71
\reference Mao, S., \& Paczy\'nski, B. 1991, ApJ, 374, L37
\reference Nemiroff, R. J., \& Wickramasinghe, W. A. D. T. 1994, ApJ, 
424, L21 
\reference Paczy\'nski, B. 1986, ApJ, 304, 1 
\reference   Udalski, A., Szyma\'nski, M., Ka\l u\.zny, J., Kubiak, M., 
Krzemi\'nski, W., Mateo, M., Preston, G. W., \& Paczy\'nski, B. 
1993, Acta Astron., 43, 289 
\reference Udalski, A., Szyma\'nski, M., Stanek, K. Z., Ka\l u\.zny, J., 
Kubiak, M., Mateo, M., Krzemi\'nski, W., Paczy\'nski, B., \& 
Venkat, R. 1994a, Acta Astron., 44, 165 
\reference Udalski, A., Szyma\'nski, M., Ka\l u\.zny, J., Kubiak, M., 
Mateo, M., Krzemi\'nski, W., \& Paczy\'nski, B.  1994b, Acta Astron., 44, 227 
\reference Witt, H. J., 1995, ApJ, 449, 42
\reference Witt, H. J., \& Mao, S. 1994, ApJ, 430, 505
 
\end{references}
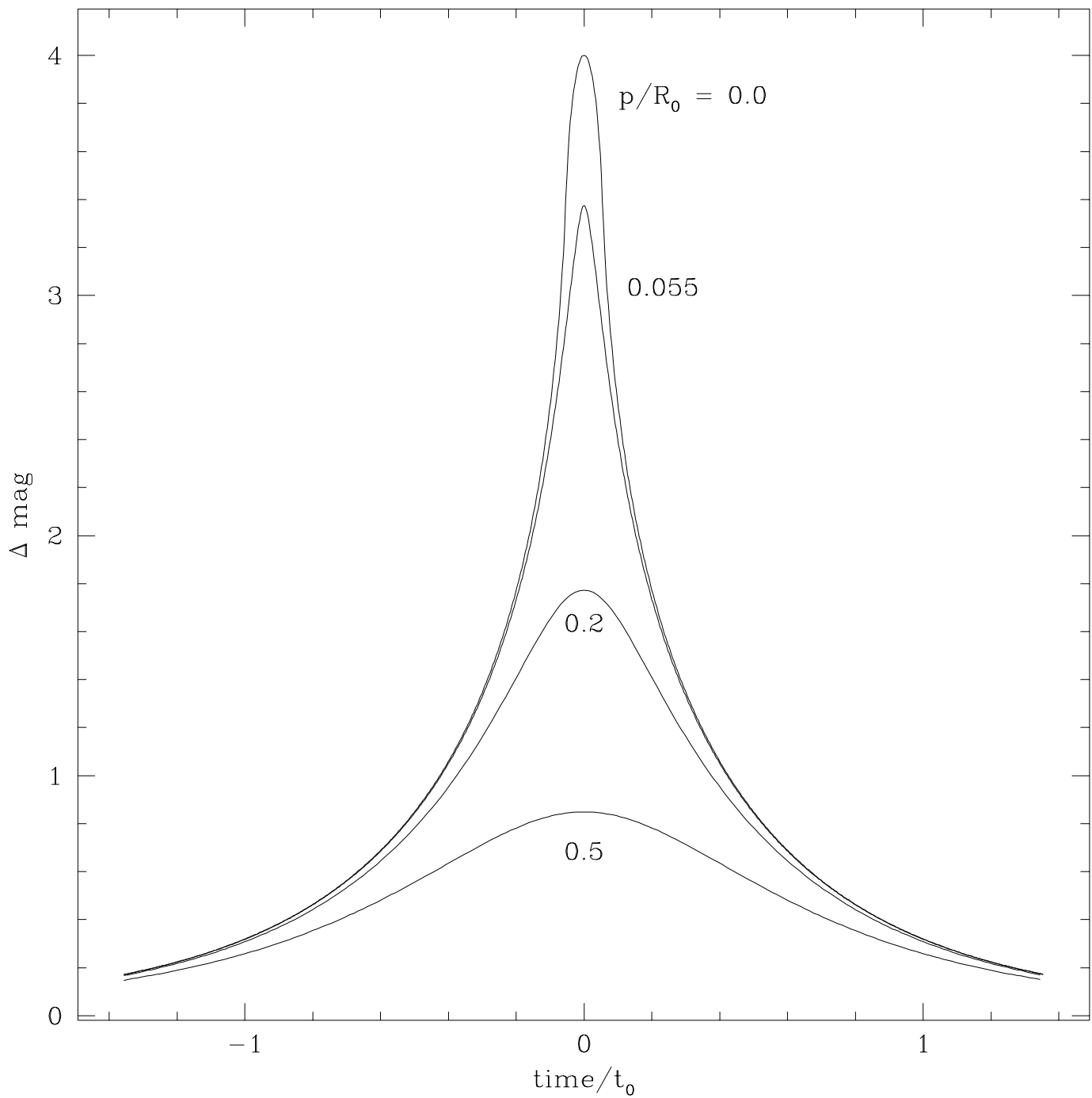
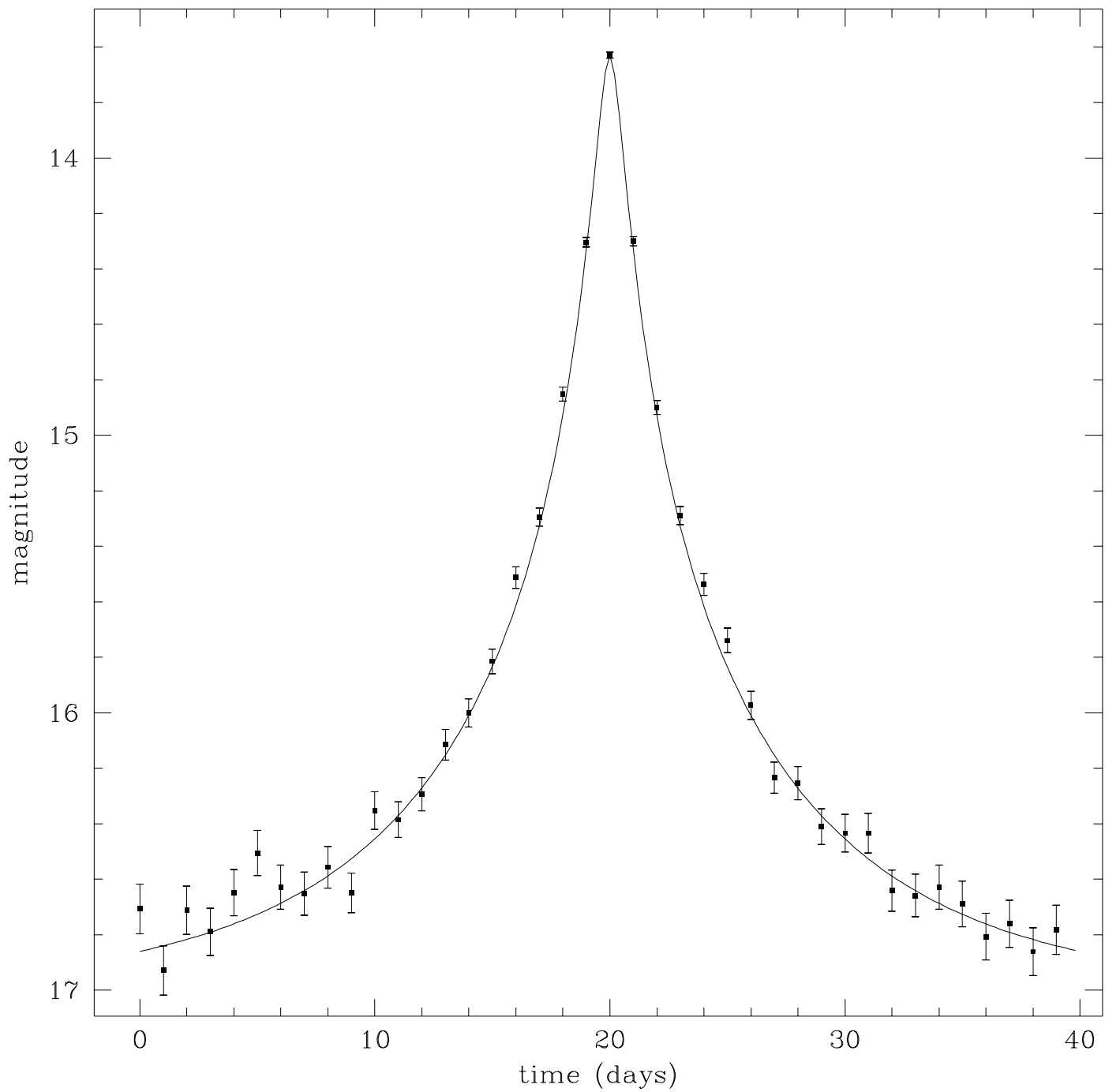
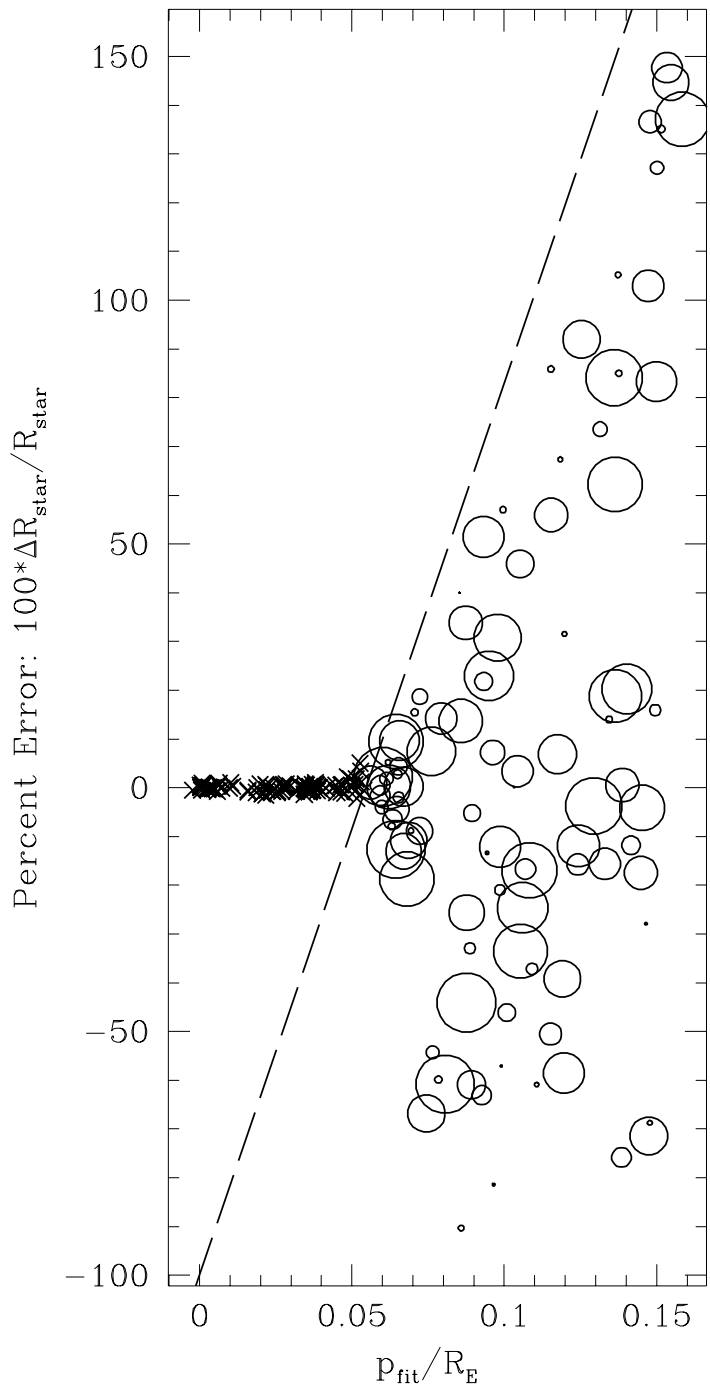
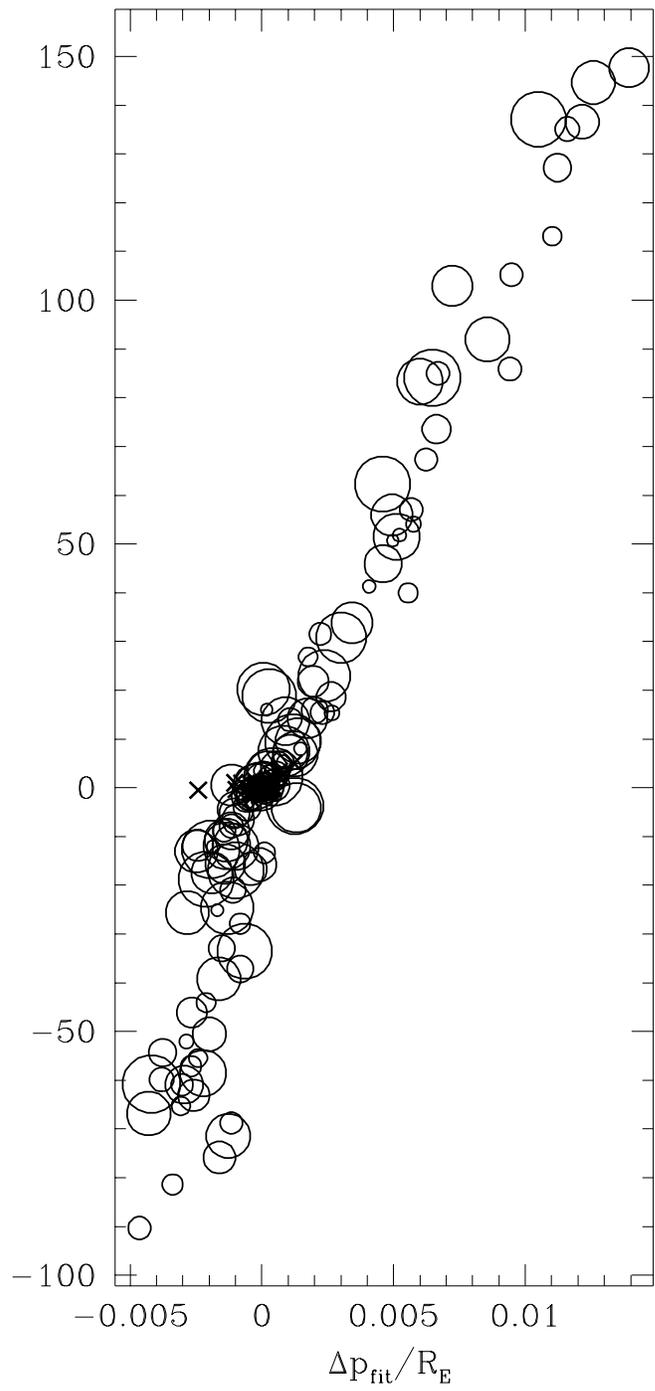
\end{document}